\documentclass{camera}
\newcommand{\be}{\begin{equation}}
\newcommand{\ee}{\end{equation}}
\newcommand{\bea}{\begin{eqnarray}}
\newcommand{\eea}{\end{eqnarray}}

\begin{document}
\title{ON INCLUSIVE $B$   DECAYS }
\author{Fulvia De Fazio}

%
\organization{Istituto Nazionale di Fisica Nucleare - Sezione di
Bari - Italy}

\maketitle

\abstract{I briefly review the description  of inclusive   heavy
hadron decays based on the   inverse heavy quark mass expansion.
In particular, I consider  the transition $B \to X_u \ell \nu$ and
the problem of  extracting  $V_{ub}$, mainly as far as  the
separation from the background represented by $b \to c$ modes is
concerned. I also discuss how to obtain complementary information
from the decay  mode $B \to X_s \gamma$.}
\section{General Framework}
Inclusive  decay widths of  hadrons $H_Q$ with a single  heavy
quark $Q$ can be computed  using  an expansion in powers of
$m_Q^{-1}$ \cite{misha}, starting from  the optical theorem that
allows to write $ \Gamma(H_Q \to X_f)=2 Im\langle H_Q |{\hat
T}|H_Q\rangle / 2 M_{H_Q} $
  with ${\hat T}=i \int d^4x T[{\cal
H}_w(x){\cal H}_w^{\dag}(0)]$ the transition operator describing
the heavy quark $Q$ with the same momentum in the initial and
final state, and ${\cal H}_w$ the effective hamiltonian governing
the decay $Q \to X_f$. An operator product expansion (OPE) of
$\hat T$ in the inverse mass of the heavy quark: ${\hat T}=\sum_i
C_i {\cal O}_i$ can be given in terms of  local operators ${\cal
O}_i$ ordered by increasing dimension, and
 coefficients $C_i$ proportional to increasing powers of $m_Q^{-1}$.
As a result, for a beauty hadron $H_b$ the general expression of
the inclusive width $\Gamma(H_b \to X_f)$ is:
\begin{equation}
 \Gamma(H_b \to X_f)=\Gamma_0 \Big[c_3^f \langle {\bar b}b
\rangle_{H_b} +{c_5^f \over m_b^2} \langle {\bar b} i g_s \sigma
\cdot G b \rangle_{H_b} +{\cal O}\Big({1 \over m_b^3} \Big) \Big]
\;\; , \label{ris}
\end{equation}
with $\displaystyle{\langle O \rangle_{H_b}={\langle
H_b|O|H_b\rangle \over 2 M_{H_b}}}$,
$\displaystyle{\Gamma_0={G_F^2 m_b^5 \over 192 \pi^3}|V_{qb}|^2}$
and $V_{qb}$ the relevant CKM  matrix element.

The first operator in (\ref{ris}) is ${\bar b}b$, with dimension
$D=3$; the chromomagnetic operator ${\cal O}_G={\bar b}{g \over 2}
\sigma_{\mu \nu} G^{\mu \nu}b$, responsible of the heavy
quark-spin symmetry breaking, has $D=5$. In the limit $m_b \to
\infty$, the heavy quark equation of motion allows to write:
\begin{equation}
\langle {\bar b}b\rangle_{H_b}= 1+{\langle {\cal O}_G
\rangle_{H_b} \over 2 m_b^2} -{ \langle{\cal O}_\pi \rangle_{H_b}
\over 2 m_b^2} +{\cal O}\Big({1 \over m_b^3}\Big) , \label{bbarb}
\end{equation}
with ${\cal O}_\pi={\bar b}(i {\vec D})^2b$ the heavy quark
kinetic energy operator. Inserting (\ref{bbarb}) in (\ref{ris}),
the leading term
 reproduces the spectator model result. ${\cal
O}(m_b^{-1})$ terms  are absent \cite{Chay:1990da} since $D=4$
operators are reducible to  ${\bar b}b$ by the equation of motion.
Finally, the operators ${\cal O}_G$ and ${\cal O}_\pi$ are
spectator blind, not sensitive to light flavour. Their matrix
elements $ \mu_G^2(H_b)=\langle {\cal O}_G \rangle_{H_b} $, $
\mu_\pi^2(H_b)=\langle {\cal O}_\pi \rangle_{H_b} $  enter in the
mass formula
 \be
M_{H_b}=m_b+{\bar \Lambda}+{\mu_\pi^2 -\mu_G^2 \over 2 m_b} +{\cal
O}({m_b^{-2}}) \ee and in principle can  be determined
experimentally. $\mu^2_G$ can be  obtained  from the  mass
splitting; for $B$ mesons:
 $ \mu_G^2(B)=3(M_{B^*}^2-M_B^2)/4$.
${\cal O}(m_b^{-3})$ terms in (\ref{ris}) come from four-quark
operators and are responsible for lifetime differences between
baryons and mesons.

Each term in (\ref{ris}) can be further  expanded in powers of
$\alpha_s$, therefore one envisages in this  approach a model
independent framework to compute inclusive decay widths and to
reliably  determine $V_{ub}$ or $V_{cb}$. The drawback is that
predictions are really reliable only if truly inclusive quantities
are considered, while it is often necessary to put cuts on the
relevant kinematical variables. A further uncertainty is
represented by quantities such as $m_b$ or $\bar \Lambda$, which
enter as input parameters, although one aims at determining them
from data, as well.

Finally, an open question remains the problem of the lifetime
ratio $\tau(\Lambda_b) /\tau(B_d)$, where  it  seems still
unlikely  that the inclusion of ${\cal O}(m_b^{-3})$ terms allows
to reproduce the current experimental datum
\cite{Colangelo:1996ta}.
\section{Inclusive $B \to X_u \ell \nu$ transition}

In the  application  of the previous formalism to the  decay $B
\to X_u \ell \nu$ for the extraction of $V_{ub}$  the main problem
is represented by the background due to $b \to c $ transitions.
One should consider differential distributions in the relevant
kinematical variables
 and put suitable cuts in order to
achieve  an effective discrimination. However, in some of the
resulting regions of the phase space the OPE is no more reliable,
and singular terms appear that signal the inadeguacy of the
approach.

 Let us
consider the distribution in the lepton energy $E_\ell$; one can
subtract the charm background imposing the cut  $E_\ell > (m_B^2
-m_D^2)/ 2 m_B$ that allows only the lightest (charmless)
particles to be produced. Since the largest energy available is
$E_\ell^{max}=(m_B^2 -s_H )/2 m_B$, where $s_H$ is the hadronic
invariant mass,  the range which can be  experimentally useful is
$(\Delta E_\ell)_{end-point}=m_D^2/2 m_B \simeq 0.33 \,\, GeV$,
 too small to guarantee a significant statistics. In this region
 the OPE is also no more reliable, since the true
expansion parameter is ${\bar \Lambda}/(m_b -2 E_\ell)$, which is
large in the end point.

Actually, the calculation of the lepton energy spectrum shows the
appearance of singular distributions $\delta^{(n)}(y-1)$, where
$y=2 E_\ell/m_b$. The inadeguacy of the approach is also evident
from the fact that although $m_b/2$ is the largest energy
available for a free quark decay, the true end point corresponds
to  $E_\ell =m_B /2$. In this {\it window} bound state effects,
due to the Fermi motion of the heavy quark, become important. They
can be taken into account  introducing a non perturbative form
factor, known as shape function, representing a resummation of all
the singular terms \cite{sf}. The physical spectra  are then
obtained through a convolution of the differential distributions
with such a function. The knowledge of the first terms of the
expansion of the shape function, gives us information about its
first few moments. However, the reconstruction procedure is not
unique and the uncertainty linked to the shape function hampers
the extraction of $V_{ub}$.

The hadronic invariant mass distribution is more promising.
 To subtract the charm background one should impose $s_H
<m_D^2$ and, in order to assess the effectiveness of considering
such a distribution, it is necessary to evaluate how many events
are left after the cut is imposed: \be \Gamma(m^2)=\int_{{\bar
\Lambda}^2}^{m^2} ds_H {d \Gamma \over ds_H} \hskip 1.5 cm {\bar
\Lambda}^2 \le {m^2}\le m_B^2 \,.\label{spec-int} \ee In
\cite{DeFazio:1999sv} the fully differential distribution relative
to the decay $B \to X_u \ell \nu$ has been calculated including
${\cal O}(\alpha_s)$ corrections, reducing the theoretical
uncertainty linked to such corrections. The results for the
hadronic invariant mass spectrum show that the cut $s_H <m_D^2$
reduces only slightly the number of events, while eliminating
completely the charm background. In order to consider the
realistic situation, in \cite{DeFazio:1999sv} the spectrum at
${\cal O}(\alpha_s)$ has been convoluted with a form of the shape
function  \cite{Kagan:1998ym}, obtaining that the number of events
below the cut $s_H <m_D^2$ is $80 \pm 10 \%$. Therefore, the
analysis of the hadronic invariant mass distribution should be
effective for the determination of $V_{ub}$.

An alternative procedure is to consider
 the spectrum in the leptonic invariant mass $q^2$ \cite{Bauer:2000xf}.
  The cut
$q^2>(m_B-m_D)^2$ discriminates the charm signal. The range where
the OPE is no more reliable can  roughly be estimated as the
window between the parton end-point $q^2=m_b^2$, and the true
end-point $q^2=m_B^2$. However, the range that can be exploited
experimentally is much larger: $\Delta q^2=m_B^2-(m_B-m_D)^2$,
indicating that non perturbative effects should have a minor
impact in this case. Nevertheless, the fraction of events
surviving to the cut is only $\simeq 20 \%$. Probably, from the
experimental side, the most suitable procedure is the analysis of
a double differential distribution, e.g. both in $s_H$ and $q^2$
\cite{Bauer:2001rc}.

An important step towards the reduction of the theoretical
uncertainty in the extraction of $V_{ub}$ from  $B \to X_u \ell
\nu$ comes from the analysis of the photon energy spectrum in $B
\to X_s \gamma$. In the quark transition $b \to s \gamma$ such a
spectrum would be monochromatic. However, real gluon emission as
well as the Fermi motion modify the spectrum. The effect of the
Fermi motion can be included introducing the same form factor as
in $B \to X_u \ell \nu$ \cite{sf}. Since real gluon emission is
computable in perturbation theory, one could try to extract
information about the shape function from this mode. This is what
the CLEO Collaboration is  doing, and, from the combined analysis
of the modes $B \to X_s \gamma$ and $B \to X_c \ell \nu$, the
first moments of the shape function have been determined, leading
to the results \cite{Chen:2001fj}: \be \mu_\pi^2 =0.236 \pm 0.071
\pm 0.078 \,GeV^2 \hskip 0.8 cm \bar \Lambda = 0.35 \pm 0.08 \pm
0.10 \, GeV \label{cleo1}\,.\ee The experimental analysis is
ongoing, as well as the theoretical improvements, reinforcing  our
hope to get soon more precise information.

{\bf Acknowledgments} I thank P. Colangelo and M. Neubert for
collaboration on some of the topics discussed above.

\end{document}